\documentclass[aip,jap,reprint]{revtex4-1}

\usepackage{graphicx}% Include figure files
\begin{document}
%\linenumbers
%\preprint{}

\title{Growth and characterization of multiferroic BiMnO$_3$ thin films} %Title of paper



\author{Hyoungjeen Jeen}
%\affiliation{Department of Physics, University of Florida, Gainesville, Florida 32611, USA}
\author{Guneeta Singh-Bhalla}
\altaffiliation[Presently at: ]{Department of Physics, University of California, Berkeley, California 94720, USA 
and Materials Science Division, Lawrence Berkeley National
Laboratory, Berkeley, California 94720, USA}
%\affiliation{Department of Physics, University of Florida, Gainesville, Florida 32611, USA}
%\affiliation{Department of Physics, University of California, Berkeley, California 94720, USA}

\author{Patrick R. Mickel}
%\affiliation{Department of Physics, University of Florida, Gainesville, Florida 32611, USA}
\author{Kristen Voigt}
%\affiliation{Department of Physics, University of Florida, Gainesville, Florida 32611, USA}
\author{Chelsey Morien}
%\affiliation{Department of Physics, University of Florida, Gainesville, Florida 32611, USA}
\author{Sefaattin Tongay}
\altaffiliation[Also at: ]{Nanoscience Institute of Medical and Engineering Technology, University of Florida, 32611}
%\affiliation{Department of Physics, University of Florida, Gainesville, Florida 32611, USA}
\author{A.~F.~Hebard}
%\affiliation{Department of Physics, University of Florida, Gainesville, Florida 32611, USA}
\author{Amlan Biswas}
\email{amlan@phys.ufl.edu}
\affiliation{Department of Physics, University of Florida, Gainesville, Florida 32611, USA}


\date{\today}

\begin{abstract}
We have grown epitaxial thin films of multiferroic BiMnO$_3$ using pulsed laser deposition. The films were grown on SrTiO$_3$ (001) substrates by ablating a Bi-rich target. Using x-ray diffraction we confirmed that the films were epitaxial and the stoichiometry of the films was confirmed using Auger electron spectroscopy. The films have a ferromagnetic Curie temperature ($T_C$) of 85$\pm$5 K and a saturation magnetization of 1 $\mu_B$/Mn. The electric polarization as a function of electric field ($P-E$) was measured using an interdigital capacitance geometry. The $P-E$ plot shows a clear hysteresis that confirms the multiferroic nature of the thin films.
\end{abstract}

\pacs{Valid pacs appear here}% insert suggested PACS numbers in braces on next line

\maketitle %\maketitle must follow title, authors, abstract and \pacs
Multiferroic materials are unique in that they exhibit both ferromagnetism and ferroelectricity simultaneously. \cite{RefWorks:77} Such materials may be used to fabricate devices such as magnetic tunnel junctions with electrically tunable tunneling magnetoresistance and multiple state memory elements.\cite{RefWorks:34} The recent interest in multiferroics is fueled both by the potential device applications and questions about the underlying physical principles leading to multiferroism.\cite{RefWorks:58, RefWorks:49, RefWorks:24, RefWorks:20, RefWorks:47} Bulk multiferroic materials are rare, possibly due to conflicting requirements for ferromagnetism (FM) and ferroelectricity (FE). BiMnO$_3$ is perhaps the most fundamental multiferroic and has been referred to as the ``hydrogen atom'' of multiferroics.\cite{Refworks:245} In BiMnO$_3$ (BMO), as in BiFeO$_3$, the 6s$^2$ lone pair on the Bi-ion leads to the displacement of that ion from the centrosymmetric position at the A-site of a perovskite unit cell. The resultant distortion leads to an FM interaction between the Mn-ions at the B-site in BMO.\cite{RefWorks:12, RefWorks:17} In bulk form BMO has been observed to be both FM and FE.\cite{RefWorks:65} Polycrystalline BMO can be grown under high pressure and within a very narrow range of growth conditions. While thin films of BMO have been grown by various groups, few such films have shown magnetic properties similar to bulk BMO and high enough resistivities i.e. low leakage currents to allow clear measurement of FE properties.\cite{RefWorks:214,RefWorks:257,RefWorks:258} A possible reason for the low resistivities of BMO thin films is the substrate induced strain which exacerbates the growth of a highly distorted perovskite structure. Additionally, recent electron and neutron diffraction data have cast doubt over the purported non-centrosymmetry of the BMO crystal structure\cite{RefWorks:252} and centrosymmetric structures have also been predicted using density functional theory calculations\cite{RefWorks:251}. Since a non-centrosymmetric crystal structure is essential for ferroelectricity, the observed ferroelectric behavior of BMO may be due to strain and/or ordered oxygen vacancies.\cite{RefWorks:249,RefWorks:255}

\begin{figure}[b]
\includegraphics[width=8cm,height=8.5cm]{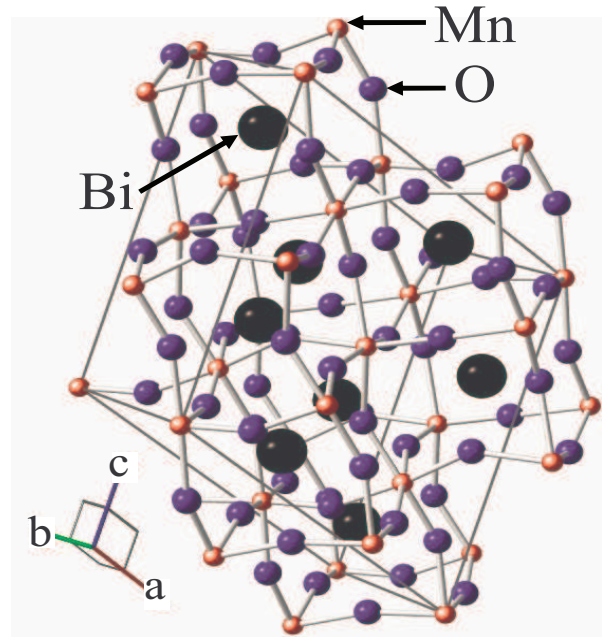}\\
\caption{The monoclinic unit cell of BiMnO$_3$.\cite{Refworks:256}}
\end{figure}

BMO has a distorted perovskite-type structure with $a = c = 3.935$ \AA~ ($\alpha = \gamma = 91.4^{\circ}$) and $b = 3.989$ \AA~ ($\beta = 91^{\circ}$).\cite{RefWorks:66} Fig. 1 shows the larger monoclinic unit cell of BMO\cite{RefWorks:12}; we have used the monoclinic notation to index the x-ray diffraction data of our thin films. Since cubic SrTiO$_3$ (STO) has a lattice parameter of 3.905 \AA, BMO grows with an (111) orientation on STO (001) substrates under a compressive strain due to a lattice mismatch of 0.77\%. It is still unclear whether this strain is responsible for the difference in the magnetic and electrical properties between thin films and polycrystalline BMO. A saturation magnetization $M_{sat}$ of about 3.6 $\mu_B$/Mn at 5 K and a ferromagnetic Curie temperature ($T_C$) of 105 K has been observed in polycrystalline BMO along with an electric remnant polarization of 62 nC/cm$^2$ at 87 K, while in thin films an $M_{sat}$ of about 2.2 $\mu_B$/Mn and a $T_C$ of about 100 K has been reported.\cite{RefWorks:65,RefWorks:66,RefWorks:214,RefWorks:227} $P-E$ measurements on BMO thin films have been reported occasionally, and a remnant polarization of about 16 $\mu$C/cm$^2$ has been observed.\cite{RefWorks:258} The strain may also influence the ferroelectric domain wall motion which coupled with the low resistance of the thin films have made it a challenge to confirm the FE nature of BMO thin films leading to the controversial situation presented in the introduction. To address such issues, we have optimized the growth of BMO on STO. We have obtained stoichiometric, epitaxial thin films of BMO which have a high resistivity at low temperature thus facilitating the measurement of $P-E$ loops confirming the multiferroic nature of our films.

The BMO thin films were grown using pulsed laser deposition (PLD). An off-stoichiometric (Bi-rich) target with composition Bi$_{2.4}$MnO$_3$ was ablated using a KrF excimer laser ($\lambda$ = 248 nm). The high Bi content of the target allowed us to use relatively high substrate temperatures ($T_s$) and still get the right Bi content for stoichiometric BMO films. The film quality was extremely sensitive to the $T_s$ and the oxygen pressure, and only slightly sensitive to the laser energy, while it was independent of the growth rate within the range used. The laser energy density was kept at 1.0 $\pm$ 0.2 J/cm$^2$. The optimum flowing oxygen pressure and $T_s$ were 37 mTorr and 632$^{\circ}$C, respectively. The film thickness was varied from 30 nm to 60 nm and the deposition rate was 0.05 nm/s. The films were cooled in an O$_2$ atmosphere of 680 Torr at a rate of 20$^{\circ}$C/min. The surface of the films was imaged using the tapping mode in a Digital Instruments Nanoscope III atomic force microscope (AFM). The structural and chemical properties of the films were characterized with standard $\theta - 2\theta$ x-ray diffraction using a Philips APD 3720 system and Auger electron spectroscopy (AES) using a Perkin-Elmer PHI 660 scanning Auger multiprobe instrument. The magnetization was measured using a Quantum Design SQUID magnetometer. We also measured the electrical polarization using an interdigital capacitance geometry and a Precision LC ferroelectric tester from Radiant Technologies. Here, we present results from one 60 nm-thick BMO thin film. We obtained similar results from the other thin films in the thickness range of 30 nm to 60 nm. 

\begin{figure}[t]
\includegraphics[width=8cm,height=14cm]{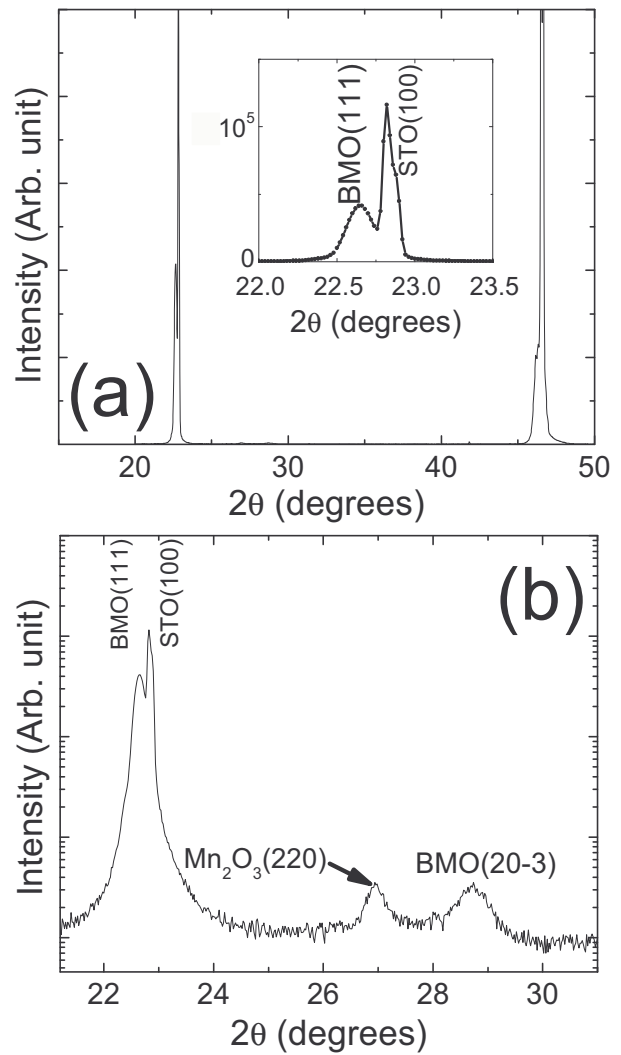}\\
\caption{(a) $\theta-2\theta$ x-ray diffraction pattern of an 60 nm-thick BiMnO$_3$ thin film. The inset shows the BiMnO$_3$ (111) peak in detail. (b) A semilog plot showing a small amount of Mn$_2$O$_3$ impurity phase and a small BMO(20$\overline{3}$) peak.}
\end{figure}

Fig. 2(a) shows the x-ray diffraction data for a 60 nm-thick BMO thin film. The inset shows that the BMO grows with a (111) orientation as expected from the structure of BMO. We also observed a small peak corresponding to Mn$_2$O$_3$ impurities which is visible in the semilog plot (Fig. 2(b)) (integrated intensity ratio of the Mn$_2$O$_3$ peak to the BiMnO$_3$ (111) peak is 0.025.). To confirm the stoichiometry of the samples we performed Auger electron spectroscopy (AES) measurements at 300K in ultra high vacuum (UHV) conditions. Derivative AES surface spectra were taken using 5 keV primary electron beam from kinetic energies of 50 eV to 1500 eV at incident angles from 30$^\circ$ to 60$^\circ$. Depth profiling was performed by taking surface spectra with the parameters given above followed by an {\em in-situ} repeated 3 keV Ar-ion sputtering. Surface spectra of the BMO films displayed three manganese (Mn) peaks located at 548 eV, 595 eV, 645 eV,  two bismuth (Bi) peaks at 106 eV, 254 eV and one oxygen (O) peak at 518 eV together with residue carbon (C) peak at 273 eV with concentrations less than ~1\%. After six seconds of Ar sputtering on the surface, the C peak disappeared and Bi, Mn and O concentrations are found to be 23.3\%, 24.1\% and 52.6\% respectively with about a 2\% error. These concentrations imply that the BiMnO$_3$ stoichiometry is consistent with the measured BMO x-ray peaks from $\theta - 2\theta$ measurements. Moreover, the sensitivity factor for oxygen is based on an MgO matrix and since there is no matrix parameter in the atomic percentage calculations, this could account for the slightly lower than stoichiometric oxygen concentrations.

\begin{figure}[t]
\includegraphics[width=8cm,height=12cm]{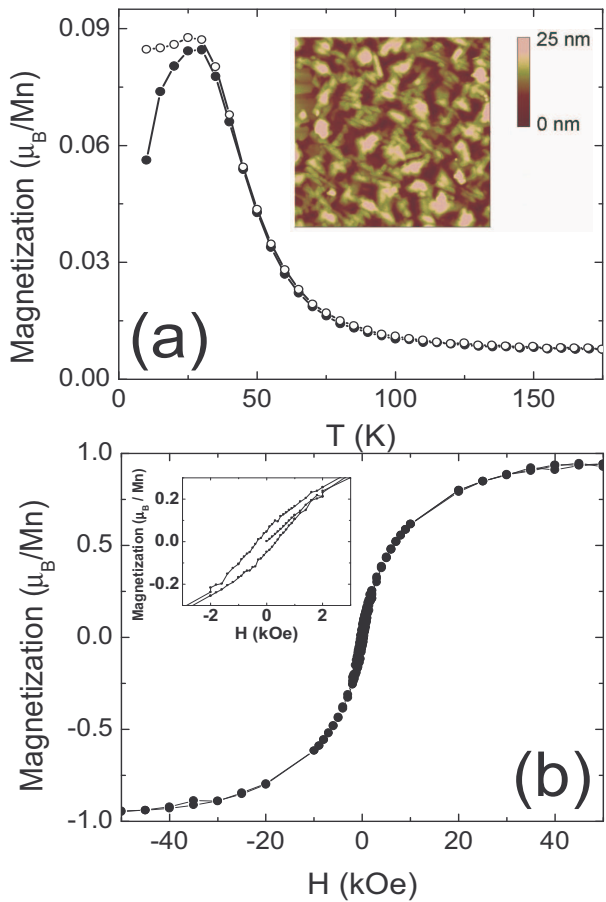}\\
\caption{(a) Magnetization vs. temperature plot for a 60 nm-thick BiMnO$_3$ thin film in an in-plane field of 500 Oe. The full circles and open circles are the zero field cooled and field cooled data respectively. The inset shows a 2 $\mu$m $\times$ 2 $\mu$m atomic force microscope image of the thin film surface. (b) Magnetization vs. magnetic field ($M - H$) plot for the 60 nm film at 10 K. The inset shows the hysteresis in the $M - H$ data.}
\end{figure}

The magnetic properties of BMO are closely related to its unique crystal structure. BMO is similar to the compound LaMnO$_3$ (LMO) but due to the 6$s$ lone pair the Bi ion moves away from the centrosymmetric position at the B-site of a perovskite structure. LMO is an A-type antiferromagnet due to antiferromagnetically stacked ferromagnetic layers.\cite{RefWorks:64} In BMO the distortion caused by the Bi-ion leads to an FM interaction between the layers.\cite{RefWorks:12, RefWorks:17} Hence, BMO has an overall magnetic moment that has been measured to be as high as 3.6 $\mu_B$/Mn in polycrystalline samples, which is close to maximum possible magnetization of 4 $\mu_B$/Mn.\cite{RefWorks:66} In thin films the magnetic moment is reduced quite likely due to the substrate induced strain. The $T_C$ in thin films is also lower than the value of about 105 K obtained in polycrystalline samples.\cite{RefWorks:65, RefWorks:66} Fig. 3 shows the $M-T$ and $M-H$ curves of a 60 nm BMO thin film. The magnetic field was applied in the plane of the film for the magnetic measurements. The $M-T$ plot reveals a $T_C$ of about 85 $\pm$ 5 K. A saturation magnetization of about 1 $\mu_B$/Mn is obtained at 10 K in a field of 50 kOe. The inset of Fig. 3(b) shows the hysteresis in the $M-H$ plot and a coercive field of about 344 Oe. The inset of Fig. 3(a) also shows the surface morphology of the thin film. The r.m.s roughness of the film was 4.7 nm

\begin{figure}
\includegraphics[width=8cm,height=9cm]{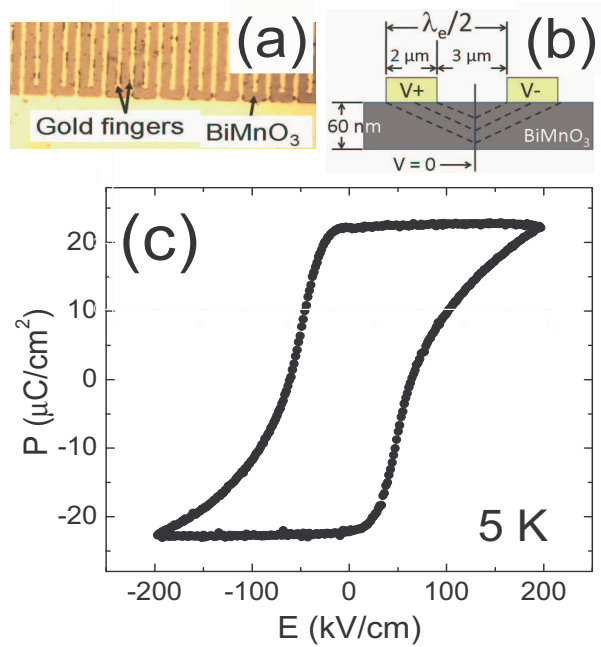}\\
\caption{(a) The interdigital capacitance geometry deposited on the film surface. (b) Schematic of the electrode configuration for the $P - E$ measurements. (c) Remnant polarization vs. electric field ($P - E$) data of a 60 nm-thick BiMnO$_3$ thin film taken at 5 K.}
\end{figure}
% \begin{figure}
% \end{figure}
A low leakage current and hence high resistivity is a requirement for polarization vs. electric field ($P - E$) measurements in BMO thin films. Our optimized thin films have a room temperature resistivity of about 10 $\Omega$-cm, which is lower than values reported by other groups.\cite{RefWorks:20,RefWorks:257} However, by 140 K (below 140 K the resistance is too high to measure with our instrumentation) the resistivity increases to about 1 M$\Omega$-cm and it was possible to make direct polarization vs. electric field ($P - E$) measurements at temperatures below 100 K using an interdigital capacitance geometry (Fig. 4(a)).  The capacitor is composed of alternating $V+$/$V-$ electrodes uniformly spaced on the film surface (Fig. 4(b)).  This structure leads to equipotential planes intersecting the film between each pair of electrodes, resulting in a capacitance between the projected areas of each electrode within the film. The projected areas were calculated using conformal mapping and equating the capacitor thickness to half the electrode spatial wavelength ($\lambda_e$ = 10$\mu$m).\cite{RefWorks:9} Fig. 4(c) shows the remnant hysteresis loop at 5 K. The polarization in a hysteresis loop is calculated by integrating the total transferred charge during application of a bipolar triangular voltage waveform.  This polarization includes contributions from leakage current, capacitance and ferroelectric domain switching. The polarization in the remnant hysteresis loop is calculated by isolating the transferred charge from only the domain-switching.  This is done by subtracting two hysteresis loops that are preceded by poling pulses.  In one loop, all of the domains are pre-switched so that no domain-switching charge is transferred during the loop, and in the other loop all the domains are set unswitched with charge transfer from domain-switching beginning at the coercive field.  The leakage current and capacitance contributions from the two loops cancel, leaving only transferred charge from domain-switching, or equivalently, the remnant charge. Dividing the remnant charge by the projected area we find a remnant polarization of $P \approx$ 23 $\mu$C/cm$^2$ at 5K, with a coercive field $E_{C} \approx $ 60 kV/cm. 

The clear observation of a ferroelectric $P-E$ loop appears to be in conflict with the centrosymmetric structure suggested by Belik {\em et al.}.\cite{RefWorks:252} If the crystal structure of BMO is indeed centrosymmetric, then the possible reasons for the ferroelectric behavior of BMO thin films are: (1) structural distortions due to oxygen vacancies,\cite{RefWorks:252} (2) a centrosymmetric to non-centrosymmetric transition below $T_C$ i.e. below 100 K,\cite{RefWorks:252} and (3) substrate induced strain.\cite{RefWorks:249} Although, the AES measurements on our thin films reveal an oxygen deficiency which could lead to the ferroelectric behavior, we cannot rule out the role of substrate induced strain. If the film is uniformly strained, the lattice mismatch which is -0.77\% (compressive), is not enough to break the centrosymmetry as shown by Hatt {\em et al.}.\cite{RefWorks:249} However, it has been shown that compressive lattice mismatch strain could lead to a non-unifrom strain distribution in the thin film due to island formation and the strain at the island edges could far exceed the average lattice mismatch strain.\cite{RefWorks:253,RefWorks:254} The growth morphology of our thin films (inset Fig. 3(a)) suggests that such non-uniform strain distribution is also a possible mechanism for the appearance of ferroelectricity. In addition, since we measured the $P-E$ loops at 5 K, the ferroelectric behavior could be due to a structural change below $T_C$. To test this hypothesis, we are currently measuring the temperature dependence of the $P-E$ loops as the temperature is increased above $T_C$.

In summary, we have grown thin films of BiMnO$_3$ (111) on SrTiO$_3$ (001) substrates. The films have the desired structure and stoichiometry. The ferromagnetic $T_C$ is 85$\pm$5 K with a saturation magnetization of about 1~$\mu_B$/Mn at 10 K. The films have a sufficiently high resistivity at low temperatures to allow the measurement of $P-E$ loops. A remnant polarization of 23 $\mu$C/cm$^2$ was measured at 5 K with a coercive field of 60 kV/cm. Non-uniform strain distribution may be responsible for the appearance of ferroelectricity in these BMO thin films. Hence, strain dependent measurements of the magnetic and electrical properties are necessary to reveal the origin of multiferroism in BMO.\cite{RefWorks:246}

This work was supported by NSF DMR-0804452 (AB) and NSF DMR-0704240 (AFH).
%\section{}
%\label{}
%\subsection{}
%\subsubsection{}

% If in two-column mode, this environment will change to single-column format so that long equations can be displayed. 
%\begin{widetext}
%$$\mbox{put long equation here}$$
%\end{widetext}

% Figures should be put into the text as floats. 
% Use the graphics or graphicx packages (distributed with LaTeX2e).
% Put the label that you will use with \ref{} command in the braces of the \label{} command.
%
% \begin{figure}
% \end{figure}

% \begin{table}
% \end{tabular}
% \end{table}

%\begin{acknowledgments}
% Put your acknowledgments here.
%\end{acknowledgments}

%\begin{references}
%Merlin.mbs v4.21 2009-07-09.
%


\begin{thebibliography}{10}%
\makeatletter
\providecommand \@ifxundefined [1]{%
 \ifx #1\undefined \expandafter \@firstoftwo
 \else \expandafter \@secondoftwo
\fi
}%
\providecommand \@ifnum [1]{%
 \ifnum #1\expandafter \@firstoftwo
 \else \expandafter \@secondoftwo
\fi
}%
\providecommand \enquote [1]{``#1''}%
\providecommand \bibnamefont  [1]{#1}%
\providecommand \bibfnamefont [1]{#1}%
\providecommand \citenamefont [1]{#1}%
\providecommand\href[0]{\@sanitize\@href}%
\providecommand\@href[1]{\endgroup\@@startlink{#1}\endgroup\@@href}%
\providecommand\@@href[1]{#1\@@endlink}%
\providecommand \@sanitize [0]{\begingroup\catcode`\&12\catcode`\#12\relax}%
\@ifxundefined \pdfoutput {\@firstoftwo}{%
 \@ifnum{\z@=\pdfoutput}{\@firstoftwo}{\@secondoftwo}%
}{%
 \providecommand\@@startlink[1]{\leavevmode}%
 \providecommand\@@endlink[0]{}%
}{%
 \providecommand\@@startlink[1]{%
  \leavevmode
  \pdfstartlink
   attr{/Border[0 0 1 ]/H/I/C[0 1 1]}%
   user{/Subtype/Link/A<</Type/Action/S/URI/URI(#1)>>}%
  \relax
 }%
 \providecommand\@@endlink[0]{\pdfendlink}%
}%
\providecommand \url  [0]{\begingroup\@sanitize \@url }%
\providecommand \@url [1]{\endgroup\@href {#1}{\urlprefix}}%
\providecommand \urlprefix [0]{URL }%
\providecommand \Eprint[0]{\href }%
\@ifxundefined \urlstyle {%
  \providecommand \doi [1]{doi:\discretionary{}{}{}#1}%
}{%
  \providecommand \doi [0]{doi:\discretionary{}{}{}\begingroup
  \urlstyle{rm}\Url }%
}%
\providecommand \doibase [0]{http://dx.doi.org/}%
\providecommand \Doi[1]{\href{\doibase#1}}%
\providecommand \selectlanguage [0]{\@gobble}%
\providecommand \bibinfo [0]{\@secondoftwo}%
\providecommand \bibfield [0]{\@secondoftwo}%
\providecommand \translation [1]{[#1]}%
\providecommand \BibitemOpen[0]{}%
\providecommand \bibitemStop [0]{}%
\providecommand \bibitemNoStop [0]{.\EOS\space}%
\providecommand \EOS [0]{\spacefactor3000\relax}%
\providecommand \BibitemShut [1]{\csname bibitem#1\endcsname}%
%</preamble>
\bibitem{RefWorks:77}%
  \BibitemOpen
  \bibfield{author}{%
  \bibinfo {author} {\bibfnamefont{H.}~\bibnamefont{Schmid}},\ }%
  \bibfield{journal}{%
  \bibinfo {journal} {Ferroelectrics}\ }%
  \textbf{\bibinfo {volume} {162}},\ \bibinfo {pages} {317} (\bibinfo {year}
  {1994})\BibitemShut{NoStop}%
\bibitem{RefWorks:34}%
  \BibitemOpen
  \bibfield{author}{%
  \bibinfo {author} {\bibfnamefont{M.}~\bibnamefont{Gajek}}, \bibinfo {author}
  {\bibfnamefont{M.}~\bibnamefont{Bibes}}, \bibinfo {author}
  {\bibfnamefont{S.}~\bibnamefont{Fusil}}, \bibinfo {author}
  {\bibfnamefont{K.}~\bibnamefont{Bouzehouane}}, \bibinfo {author}
  {\bibfnamefont{J.}~\bibnamefont{Fontcuberta}}, \bibinfo {author}
  {\bibfnamefont{A.}~\bibnamefont{Barthelemy}},\ and\ \bibinfo {author}
  {\bibfnamefont{A.}~\bibnamefont{Fert}},\ }%
  \bibfield{journal}{%
  \bibinfo {journal} {Nat. Mater.}\ }%
  \textbf{\bibinfo {volume} {6}},\ \bibinfo {pages} {296} (\bibinfo {year}
  {2007})\BibitemShut{NoStop}%
\bibitem{RefWorks:58}%
  \BibitemOpen
  \bibfield{author}{%
  \bibinfo {author} {\bibfnamefont{N.~A.}\ \bibnamefont{Spaldin}}\ and\
  \bibinfo {author} {\bibfnamefont{M.}~\bibnamefont{Fiebig}},\ }%
  \bibfield{journal}{%
  \bibinfo {journal} {Science}\ }%
  \textbf{\bibinfo {volume} {309}},\ \bibinfo {pages} {391} (\bibinfo {year}
  {2005})\BibitemShut{NoStop}%
\bibitem{RefWorks:49}%
  \BibitemOpen
  \bibfield{author}{%
  \bibinfo {author} {\bibfnamefont{C.~N.~R.}\ \bibnamefont{Rao}}\ and\ \bibinfo
  {author} {\bibfnamefont{C.~R.}\ \bibnamefont{Serrao}},\ }%
  \bibfield{journal}{%
  \bibinfo {journal} {J. Mater. Chem.}\ }%
  \textbf{\bibinfo {volume} {17}},\ \bibinfo {pages} {4931} (\bibinfo {year}
  {2007})\BibitemShut{NoStop}%
\bibitem{RefWorks:24}%
  \BibitemOpen
  \bibfield{author}{%
  \bibinfo {author} {\bibfnamefont{M.}~\bibnamefont{Fiebig}},\ }%
  \bibfield{journal}{%
  \bibinfo {journal} {J. Phys. D: Appl. Phys.}\ }%
  \textbf{\bibinfo {volume} {38}},\ \bibinfo {pages} {R123} (\bibinfo {year}
  {2005})\BibitemShut{NoStop}%
\bibitem{RefWorks:20}%
  \BibitemOpen
  \bibfield{author}{%
  \bibinfo {author} {\bibfnamefont{W.}~\bibnamefont{Eerenstein}}, \bibinfo
  {author} {\bibfnamefont{N.~D.}\ \bibnamefont{Mathur}},\ and\ \bibinfo
  {author} {\bibfnamefont{J.~F.}\ \bibnamefont{Scott}},\ }%
  \bibfield{journal}{%
  \bibinfo {journal} {Nature}\ }%
  \textbf{\bibinfo {volume} {442}},\ \bibinfo {pages} {759} (\bibinfo {year}
  {2006})\BibitemShut{NoStop}%
\bibitem{RefWorks:47}%
  \BibitemOpen
  \bibfield{author}{%
  \bibinfo {author} {\bibfnamefont{W.}~\bibnamefont{Prellier}}, \bibinfo
  {author} {\bibfnamefont{M.~P.}\ \bibnamefont{Singh}},\ and\ \bibinfo {author}
  {\bibfnamefont{P.}~\bibnamefont{Murugavel}},\ }%
  \bibfield{journal}{%
  \bibinfo {journal} {J. Phys.: Condens. Mat.}\ }%
  \textbf{\bibinfo {volume} {17}},\ \bibinfo {pages} {7753} (\bibinfo {year}
  {2005})\BibitemShut{NoStop}%
\bibitem{Refworks:245}%
  \BibitemOpen
  \bibfield{author}{%
  \bibinfo {author} {\bibfnamefont{N.~A.}\ \bibnamefont{Hill}}\ and\ \bibinfo
  {author} {\bibfnamefont{K.~M.}\ \bibnamefont{Rabe}},\ }%
  \bibfield{journal}{%
  \bibinfo {journal} {Phys. Rev. B}\ }%
  \textbf{\bibinfo {volume} {59}},\ \bibinfo {pages} {8759} (\bibinfo {year}
  {1999})\BibitemShut{NoStop}%
\bibitem{RefWorks:12}%
  \BibitemOpen
  \bibfield{author}{%
  \bibinfo {author} {\bibfnamefont{T.}~\bibnamefont{Atou}}, \bibinfo {author}
  {\bibfnamefont{H.}~\bibnamefont{Chiba}}, \bibinfo {author}
  {\bibfnamefont{K.}~\bibnamefont{Ohoyama}}, \bibinfo {author}
  {\bibfnamefont{Y.}~\bibnamefont{Yamaguchi}},\ and\ \bibinfo {author}
  {\bibfnamefont{Y.}~\bibnamefont{Syono}},\ }%
  \bibfield{journal}{%
  \bibinfo {journal} {J. Solid State Chem.}\ }%
  \textbf{\bibinfo {volume} {145}},\ \bibinfo {pages} {639} (\bibinfo {year}
  {1999})\BibitemShut{NoStop}%
\bibitem{RefWorks:17}%
  \BibitemOpen
  \bibfield{author}{%
  \bibinfo {author} {\bibfnamefont{A.~M.}\ \bibnamefont{dos Santos}}, \bibinfo
  {author} {\bibfnamefont{A.~K.}\ \bibnamefont{Cheetham}}, \bibinfo {author}
  {\bibfnamefont{T.}~\bibnamefont{Atou}}, \bibinfo {author}
  {\bibfnamefont{Y.}~\bibnamefont{Syono}}, \bibinfo {author}
  {\bibfnamefont{Y.}~\bibnamefont{Yamaguchi}}, \bibinfo {author}
  {\bibfnamefont{K.}~\bibnamefont{Ohoyama}}, \bibinfo {author}
  {\bibfnamefont{H.}~\bibnamefont{Chiba}},\ and\ \bibinfo {author}
  {\bibfnamefont{C.~N.~R.}\ \bibnamefont{Rao}},\ }%
  \bibfield{journal}{%
  \bibinfo {journal} {Phys. Rev. B}\ }%
  \textbf{\bibinfo {volume} {66}},\ \bibinfo {pages} {064425} (\bibinfo {year}
  {2002})\BibitemShut{NoStop}%
\bibitem{RefWorks:65}%
  \BibitemOpen
  \bibfield{author}{%
  \bibinfo {author} {\bibfnamefont{A.~M.}\ \bibnamefont{dos Santos}}, \bibinfo
  {author} {\bibfnamefont{S.}~\bibnamefont{Parashar}}, \bibinfo {author}
  {\bibfnamefont{A.~R.}\ \bibnamefont{Raju}}, \bibinfo {author}
  {\bibfnamefont{Y.~S.}\ \bibnamefont{Zhao}}, \bibinfo {author}
  {\bibfnamefont{A.~K.}\ \bibnamefont{Cheetham}},\ and\ \bibinfo {author}
  {\bibfnamefont{C.~N.~R.}\ \bibnamefont{Rao}},\ }%
  \bibfield{journal}{%
  \bibinfo {journal} {Solid State Commun.}\ }%
  \textbf{\bibinfo {volume} {122}},\ \bibinfo {pages} {49} (\bibinfo {year}
  {2002})\BibitemShut{NoStop}%
\bibitem{RefWorks:258}%
  \BibitemOpen
  \bibfield{author}{%
  \bibinfo {author} {\bibfnamefont{J.~Y.}\ \bibnamefont{Son}}\ and\ \bibinfo
  {author} {\bibfnamefont{Y.-H.}\ \bibnamefont{Shin}},\ }%
  \bibfield{journal}{%
  \bibinfo {journal} {Appl. Phys. Lett.}\ }%
  \textbf{\bibinfo {volume} {93}},\ \bibinfo {pages} {062902} (\bibinfo {year}
  {2008})\BibitemShut{NoStop}%
\bibitem{RefWorks:257}%
  \BibitemOpen
  \bibfield{author}{%
  \bibinfo {author} {\bibfnamefont{M.}~\bibnamefont{Gajek}}, \bibinfo {author}
  {\bibfnamefont{M.}~\bibnamefont{Bibes}}, \bibinfo {author}
  {\bibfnamefont{A.}~\bibnamefont{Barthelemy}}, \bibinfo {author}
  {\bibfnamefont{K.}~\bibnamefont{Bouzehouane}}, \bibinfo {author}
  {\bibfnamefont{S.}~\bibnamefont{Fusil}}, \bibinfo {author}
  {\bibfnamefont{M.}~\bibnamefont{Varela}}, \bibinfo {author}
  {\bibfnamefont{J.}~\bibnamefont{Fontcuberta}},\ and\ \bibinfo {author}
  {\bibfnamefont{A.}~\bibnamefont{Fert}},\ }%
  \bibfield{journal}{%
  \bibinfo {journal} {Phys. Rev. B}\ }%
  \textbf{\bibinfo {volume} {72}},\ \bibinfo {pages} {020406} (\bibinfo {year}
  {2005})\BibitemShut{NoStop}%
\bibitem{RefWorks:214}%
  \BibitemOpen
  \bibfield{author}{%
  \bibinfo {author} {\bibfnamefont{W.}~\bibnamefont{Eerenstein}}, \bibinfo
  {author} {\bibfnamefont{F.~D.}\ \bibnamefont{Morrison}}, \bibinfo {author}
  {\bibfnamefont{J.~F.}\ \bibnamefont{Scott}},\ and\ \bibinfo {author}
  {\bibfnamefont{N.~D.}\ \bibnamefont{Mathur}},\ }%
  \bibfield{journal}{%
  \bibinfo {journal} {Appl. Phys. Lett.}\ }%
  \textbf{\bibinfo {volume} {87}},\ \bibinfo {pages} {101906} (\bibinfo {year}
  {2005})\BibitemShut{NoStop}%
\bibitem{RefWorks:252}%
  \BibitemOpen
  \bibfield{author}{%
  \bibinfo {author} {\bibfnamefont{A.~A.}\ \bibnamefont{Belik}}, \bibinfo
  {author} {\bibfnamefont{S.}~\bibnamefont{Iikubo}}, \bibinfo {author}
  {\bibfnamefont{T.}~\bibnamefont{Yokosawa}}, \bibinfo {author}
  {\bibfnamefont{K.}~\bibnamefont{Kodama}}, \bibinfo {author}
  {\bibfnamefont{N.}~\bibnamefont{Igawa}}, \bibinfo {author}
  {\bibfnamefont{S.}~\bibnamefont{Shamoto}}, \bibinfo {author}
  {\bibfnamefont{M.}~\bibnamefont{Azuma}}, \bibinfo {author}
  {\bibfnamefont{M.}~\bibnamefont{Takano}}, \bibinfo {author}
  {\bibfnamefont{K.}~\bibnamefont{Kimoto}}, \bibinfo {author}
  {\bibfnamefont{Y.}~\bibnamefont{Matsui}},\ and\ \bibinfo {author}
  {\bibfnamefont{E.}~\bibnamefont{Takayama-Muromachi}},\ }%
  \bibfield{journal}{%
  \bibinfo {journal} {J. Am. Chem. Soc.}\ }%
  \textbf{\bibinfo {volume} {129}},\ \bibinfo {pages} {971} (\bibinfo {year}
  {2007})\BibitemShut{NoStop}%
\bibitem{RefWorks:251}%
  \BibitemOpen
  \bibfield{author}{%
  \bibinfo {author} {\bibfnamefont{P.}~\bibnamefont{Baettig}}, \bibinfo
  {author} {\bibfnamefont{R.}~\bibnamefont{Seshadri}},\ and\ \bibinfo {author}
  {\bibfnamefont{N.~A.}\ \bibnamefont{Spaldin}},\ }%
  \bibfield{journal}{%
  \bibinfo {journal} {J. Am. Chem. Soc.}\ }%
  \textbf{\bibinfo {volume} {129}},\ \bibinfo {pages} {9854} (\bibinfo {year}
  {2007})\BibitemShut{NoStop}%
\bibitem{RefWorks:249}%
  \BibitemOpen
  \bibfield{author}{%
  \bibinfo {author} {\bibfnamefont{A.~J.}\ \bibnamefont{Hatt}}\ and\ \bibinfo
  {author} {\bibfnamefont{N.~A.}\ \bibnamefont{Spaldin}},\ }%
  \bibfield{journal}{%
  \bibinfo {journal} {Eur. Phys. J. B}\ }%
  \textbf{\bibinfo {volume} {71}},\ \bibinfo {pages} {435} (\bibinfo {year}
  {2009})\BibitemShut{NoStop}%
\bibitem{RefWorks:255}%
  \BibitemOpen
  \bibfield{author}{%
  \bibinfo {author} {\bibfnamefont{H.}~\bibnamefont{Yang}}, \bibinfo {author}
  {\bibfnamefont{Z.~H.}\ \bibnamefont{Chi}}, \bibinfo {author}
  {\bibfnamefont{J.~L.}\ \bibnamefont{Jiang}}, \bibinfo {author}
  {\bibfnamefont{W.~J.}\ \bibnamefont{Feng}}, \bibinfo {author}
  {\bibfnamefont{J.~F.}\ \bibnamefont{Dai}}, \bibinfo {author}
  {\bibfnamefont{C.~Q.}\ \bibnamefont{Jin}},\ and\ \bibinfo {author}
  {\bibfnamefont{R.~C.}\ \bibnamefont{Yu}},\ }%
  \bibfield{journal}{%
  \bibinfo {journal} {J. Mater. Sci.}\ }%
  \textbf{\bibinfo {volume} {43}},\ \bibinfo {pages} {3604} (\bibinfo {year}
  {2008})\BibitemShut{NoStop}%
\bibitem{Refworks:256}%
  \BibitemOpen
  \bibfield{author}{%
  \bibinfo {author} {\bibfnamefont{T.~C.}\ \bibnamefont{Ozawa}}\ and\ \bibinfo
  {author} {\bibfnamefont{S.~J.}\ \bibnamefont{Kang}},\ }%
  \bibfield{journal}{%
  \bibinfo {journal} {J. Appl. Cryst.}\ }%
  \textbf{\bibinfo {volume} {37}},\ \bibinfo {pages} {679} (\bibinfo {year}
  {2004})\BibitemShut{NoStop}%
\bibitem{RefWorks:66}%
  \BibitemOpen
  \bibfield{author}{%
  \bibinfo {author} {\bibfnamefont{H.}~\bibnamefont{Chiba}}, \bibinfo {author}
  {\bibfnamefont{T.}~\bibnamefont{Atou}},\ and\ \bibinfo {author}
  {\bibfnamefont{Y.}~\bibnamefont{Syono}},\ }%
  \bibfield{journal}{%
  \bibinfo {journal} {J. Solid State Chem.}\ }%
  \textbf{\bibinfo {volume} {132}},\ \bibinfo {pages} {139} (\bibinfo {year}
  {1997})\BibitemShut{NoStop}%
\bibitem{RefWorks:227}%
  \BibitemOpen
  \bibfield{author}{%
  \bibinfo {author} {\bibfnamefont{A.~F.~M.}\ \bibnamefont{dos Santos}},
  \bibinfo {author} {\bibfnamefont{A.~K.}\ \bibnamefont{Cheetham}}, \bibinfo
  {author} {\bibfnamefont{W.}~\bibnamefont{Tian}}, \bibinfo {author}
  {\bibfnamefont{X.~Q.}\ \bibnamefont{Pan}}, \bibinfo {author}
  {\bibfnamefont{Y.~F.}\ \bibnamefont{Jia}}, \bibinfo {author}
  {\bibfnamefont{N.~J.}\ \bibnamefont{Murphy}}, \bibinfo {author}
  {\bibfnamefont{J.}~\bibnamefont{Lettieri}},\ and\ \bibinfo {author}
  {\bibfnamefont{D.~G.}\ \bibnamefont{Schlom}},\ }%
  \bibfield{journal}{%
  \bibinfo {journal} {Appl. Phys. Lett.}\ }%
  \textbf{\bibinfo {volume} {84}},\ \bibinfo {pages} {91} (\bibinfo {year}
  {2004})\BibitemShut{NoStop}%
\bibitem{RefWorks:64}%
  \BibitemOpen
  \bibfield{author}{%
  \bibinfo {author} {\bibfnamefont{J.~B.}\ \bibnamefont{Goodenough}},\ }%
  \bibfield{journal}{%
  \bibinfo {journal} {Phys. Rev.}\ }%
  \textbf{\bibinfo {volume} {100}},\ \bibinfo {pages} {564} (\bibinfo {year}
  {1955})\BibitemShut{NoStop}%
\bibitem{RefWorks:9}%
  \BibitemOpen
  \bibfield{author}{%
  \bibinfo {author} {\bibfnamefont{R.}~\bibnamefont{Igreja}}\ and\ \bibinfo
  {author} {\bibfnamefont{C.~J.}\ \bibnamefont{Dias}},\ }%
  \bibfield{journal}{%
  \bibinfo {journal} {Sens. Actuator A- Phys.}\ }%
  \textbf{\bibinfo {volume} {112}},\ \bibinfo {pages} {291} (\bibinfo {year}
  {2004})\BibitemShut{NoStop}%
\bibitem{RefWorks:253}%
  \BibitemOpen
  \bibfield{author}{%
  \bibinfo {author} {\bibfnamefont{A.}~\bibnamefont{Biswas}}, \bibinfo {author}
  {\bibfnamefont{M.}~\bibnamefont{Rajeswari}}, \bibinfo {author}
  {\bibfnamefont{R.~C.}\ \bibnamefont{Srivastava}}, \bibinfo {author}
  {\bibfnamefont{Y.~H.}\ \bibnamefont{Li}}, \bibinfo {author}
  {\bibfnamefont{T.}~\bibnamefont{Venkatesan}}, \bibinfo {author}
  {\bibfnamefont{R.~L.}\ \bibnamefont{Greene}},\ and\ \bibinfo {author}
  {\bibfnamefont{A.~J.}\ \bibnamefont{Millis}},\ }%
  \bibfield{journal}{%
  \bibinfo {journal} {Phys. Rev. B}\ }%
  \textbf{\bibinfo {volume} {61}},\ \bibinfo {pages} {9665} (\bibinfo {year}
  {2000})\BibitemShut{NoStop}%
\bibitem{RefWorks:254}%
  \BibitemOpen
  \bibfield{author}{%
  \bibinfo {author} {\bibfnamefont{Y.}~\bibnamefont{Chen}}\ and\ \bibinfo
  {author} {\bibfnamefont{J.}~\bibnamefont{Washburn}},\ }%
  \bibfield{journal}{%
  \bibinfo {journal} {Phys. Rev. Lett.}\ }%
  \textbf{\bibinfo {volume} {77}},\ \bibinfo {pages} {4046} (\bibinfo {year}
  {1996})\BibitemShut{NoStop}%
\bibitem{RefWorks:246}%
  \BibitemOpen
  \bibfield{author}{%
  \bibinfo {author} {\bibfnamefont{J.}~\bibnamefont{Tosado}}, \bibinfo {author}
  {\bibfnamefont{T.}~\bibnamefont{Dhakal}},\ and\ \bibinfo {author}
  {\bibfnamefont{A.}~\bibnamefont{Biswas}},\ }%
  \bibfield{journal}{%
  \bibinfo {journal} {J. Phys.: Condens. Mat.}\ }%
  \textbf{\bibinfo {volume} {21}},\ \bibinfo {pages} {192203} (\bibinfo {year}
  {2009})\BibitemShut{NoStop}%
\end{thebibliography}
\end{document}